# REA, Triple-Entry Accounting and Blockchain:

## Converging Paths to Shared Ledger Systems

Juan Ignacio Ibañez[1]

Chris N. Bayer[2]

Paolo Tasca[3]

Jiahua Xu[4]

ABSTRACT

During the last half century, the concept of shared ledger systems offering a single source of truth has been repeatedly calling traditional bookkeeping into question. Improving upon the long-standing double-entry systems, solutions such as the Resource-Event-Agent (REA) accounting framework, triple-entry accounting (TEA) and blockchain have been advanced. However, to date, the historical development of shared ledger systems remains unclear and under-researched. This paper conducts a genealogical analysis of shared ledger systems, in particular tracing the development of TEA. We show how the REA framework had a certain influence over independent streams of research in the field of TEA, and how this interaction can be traced to the present incarnation of shared ledger systems in blockchain. In doing so, we duly acknowledge the influence of each individual contributing to this development, correct common misconceptions and map out how the paths of REA, TEA and blockchain overlap in the realm of shared ledger systems.

[1] DLT Science Foundation, UCL Centre for Blockchain Technologies.

[2] Development International e.V.

[3] DLT Science Foundation, UCL Centre for Blockchain Technologies.

[4] DLT Science Foundation, UCL Centre for Blockchain Technologies.

This paper also benefited from comments by a number of individuals per interview or per written correspondence. We would like to gratefully acknowledge Chris Cook, Chris Odom, Craig S. Wright, David Hartley, G. Ken Holman, Ian Grigg, Robert Haugen and Todd Boyle. Any mistakes are our own.

## I. INTRODUCTION

Scientific revolutions begin with extraordinary investigations that advance normal research (Kuhn 1996). According to Kuhn (1996, 91), this is a very special stage in paradigm shifts, in which "scientists take a different attitude toward existing paradigms, and the nature of their research changes accordingly," with a "willingness to try *anything*…"

In the accounting profession, one such set of investigations took place between 1458 and 1494, with the works of Benedetto Cotrugli (1573), Marino de Raphaeli (1475) and Luca Pacioli (1494). The result of these investigations, double-entry accounting, was applied for five centuries with widespread acceptance. However, a second revolution that took place between 1982 and 2008 has been rather overlooked.

Shared ledger systems emerged as a fruit of this revolution. One prime example of these systems is triple-entry accounting, or TEA. Although this novel accounting concept does not change the fundamental philosophy underpinning accounting, it challenges the very role of the bookkeeper by addressing the possibility where the integrity of the bookkeeper is implicated.

In recent years, TEA has begun to catch the public attention from far-related fields like computer science. Indeed, the computer science community has managed to materialize the fundamental principles of TEA through the blockchain technology (Vijai, Elayaraja, Suriyalakshimi, and Joyce 2019).[5] Meanwhile, TEA has inspired a number of works on real-time auditing and continuous assurance from the accounting community (Dai 2017; Dai and Vasarhelyi 2017; Alawadhi et al. 2015; Auditchain 2018).

These valuable and interrelated lines of investigation on TEA have not yet been duly pursued (Cai 2019, 3). The history of TEA is told in an incomplete manner, leaving the significance and implications of this scientific revolution poorly understood. Due to the gaps in the tale, TEA has been widely believed to be the result of a cryptographic revolution with radical implications for accounting (Rao 2020; see also Cai 2019; Gröblacher and Mizdraković 2019; Inghirami 2019; Pacio 2018a).

[5] In the remainder we refer to blockchain or blockchain technology in order to encompass all the possible architectural configurations and, for the sake of simplicity, also the larger family of distributed ledger technologies, i.e., community consensus-based distributed ledgers where the storage of data is not based on chains of blocks. For further understanding of the family of the distributed ledger technologies see Tasca and Tessone (2019).

However, the reverse is also true.

In this paper, we posit the existence of an opposite chain of causality: a revolution in *accounting* broadened the applicability of cryptography. Specifically, we claim that the research of William E. McCarthy (1982), Todd Boyle (2000a, 2000b, 2000c), Ian Grigg (2005b) and, possibly, also Satoshi Nakamoto (2008), constitute an interrelated set of works telling the cohesive story of shared ledger systems.

Moreover, we show that TEA is, in part, a historical byproduct of the Resource-Event-Agent (REA) accounting framework designed by McCarthy. While parallels between TEA and the REA framework have been noted (A. Gomaa et al. 2019; Grigg 2017a, 2017b, 2017c,), the historical influence of the latter over the former remains overlooked as a result of an underappreciation of the stream of work revealing the “missing link” between REA and TEA, primarily carried out by Todd Boyle (2000a, 2000b, 2000c).

With this in mind, we seek to fill in these gaps by conducting a genealogical analysis of shared ledger systems, correcting common misconceptions and investigating the largely unacknowledged influence of William E. McCarthy on TEA and today’s blockchain technology.

In order to achieve this goal, we conduct a comprehensive literature review that covers: William E. McCarthy (1982), Robert Haugen (Haugen and McCarthy, 2000), Todd Boyle (2000a, 2000b, 2000c), Ian Grigg (2005b), G. Ken Holman (2019), Yuji Ijiri (1975, 1982) and Chris Cook (2002). We place a particular emphasis on the works of Todd Boyle, the missing link between REA and TEA. We also spoke with pioneers in REA, TEA and blockchain to document the oral history of shared ledger systems. As shown by Figure 1, we find that the current explosion of shared ledger systems use cases result from the convergence of three streams of research, developing in parallel and occasionally interacting with each other.

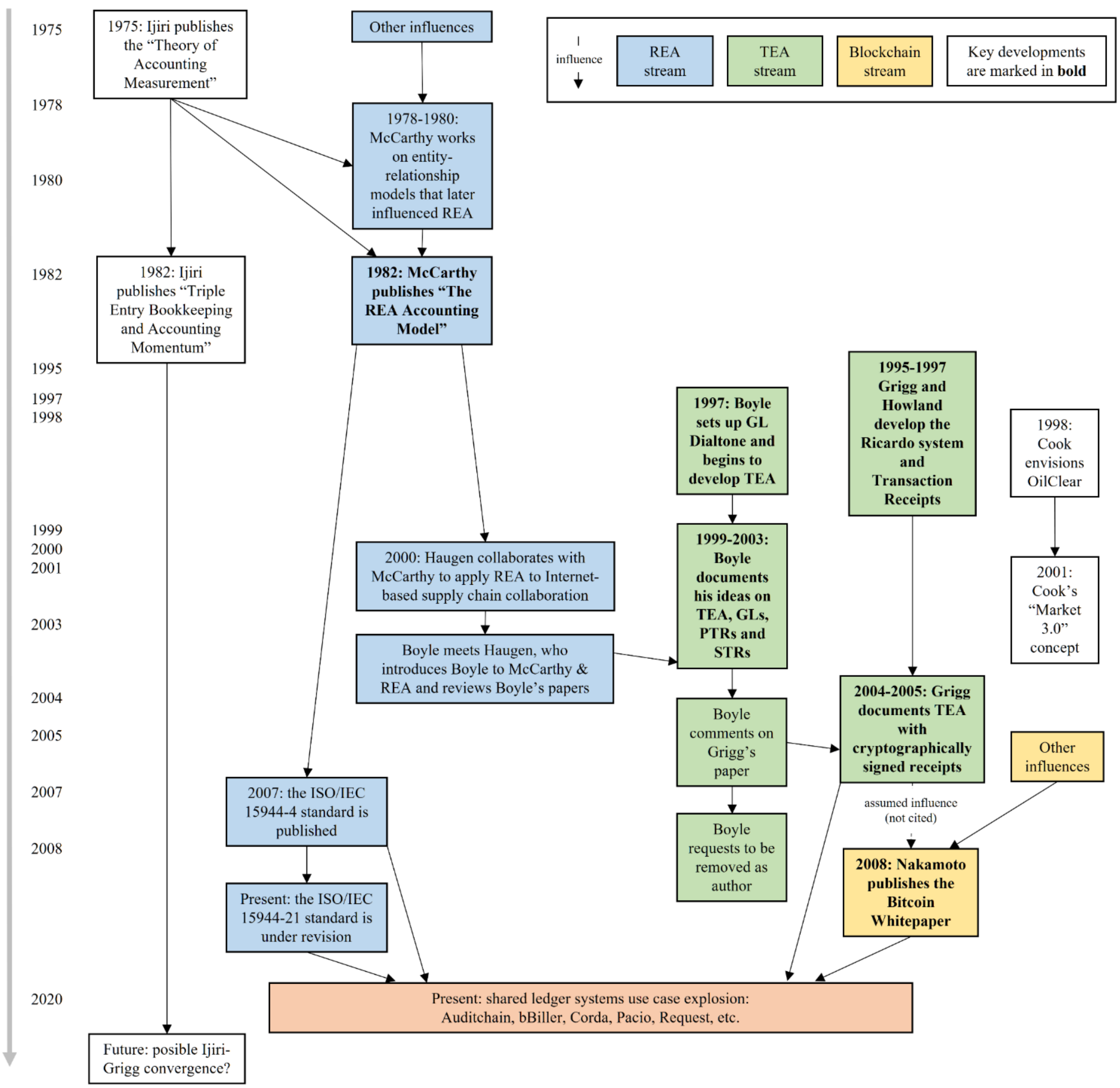

FIGURE 1: ESSENTIAL HISTORY OF TEA, SHOWING THE PARALLEL DEVELOPMENT AND CONFLUENCE OF THREE STREAMS OF RESEARCH ON SHARED LEDGER SYSTEMS, NAMELY REA, TEA AND BLOCKCHAIN.

The rest of the paper is structured in the following manner. Firstly, we specify the terminology elaborating on essential concepts of shared ledger systems. Next, we discuss their historical development. Finally, we conclude with some general considerations.

## II. CONCEPTS

### Triple-Entry Bookkeeping

Our discussion of the historical development of TEA begs the question of what TEA is in the first place. We base our taxonomy on Grigg (2019a), who distinguishes between triple-entry bookkeeping (TEB) and triple-entry accounting (TEA).

Triple-entry bookkeeping is defined as recordkeeping for two or more parties through a shared transaction repository (STR) with a "signature – signature – signature" structure (Grigg 2005a), i.e. a "triple-signed receipt" (Odom 2013, 2015). This means that, in order to update the shared entry, two parties need to be involved: one party makes a unilateral transaction entry ("request," "offer" or "transaction draft") and the other approves this entry ("acceptance"). This can be seen as a signature-gathering process: one party adds their signature to the transaction entry draft, the counterparty accepts by signing, the entry becomes valid and it is stored in the STR. The STR checks the validity of the signatures through a middleware server or distributed ledger technology network and then itself signs off on the entry (Boyle 2000e, 2001f; Grigg 2005b). This generates a hashed triple-signed receipt, such that all the parties hold the same data and that it guarantees that the data cannot be manipulated or lost: a single, shared entry, which is the *single source of truth*. In computer science terms, this is an implementation of the WYSIWIS ("What You See Is What I See") principle (Brown 2020).

In TEB, a local copy of the shared transaction repository may be integrated as a subledger to the general ledger of transactional parties. This is termed a "stub – shared entry – stub" [6] structure by Boyle (2001b). However, because the shared entry is the sole reliable source of the transaction record, some call this "single-entry bookkeeping" (Pacio 2018a, 2020). Nevertheless, we discourage usage of this term. For one, the term "single-entry bookkeeping" is already reserved for simplified bookkeeping systems in opposition to double-entry bookkeeping. These systems only recorded stock accounts, i.e. assets and liabilities, without including flow accounts such as revenues and expenses (Ijiri 1986, 746) that double-entry bookkeeping systems do, and without two entries or sides (debits and credits) for each transaction (Grigg 2005b)., Different from this sort of simplified, single-entry bookkeeping, TEB is compatible with double-entry systems (Boyle 2000c, 2000a, 2000b). In effect, TEA represents pairs of double entries with each pair connected to a central receipt, resulting in three parties holding the triple-signed receipts (Grigg 2005b). This is another reason to call this concept triple-entry, rather than single-entry. Finally, even if there is

[6] A stub is the counterfoil of a transaction receipt. As envisioned by Boyle (2001b), parties may optionally insert non-essential data in them.

a solid case for the usage the latter term, the former one is already an established category in the industry (Gröblacher and Mizdraković 2019)

At this point, we want to warn the reader that the word "entry" is polysemic. In "historical" single-entry bookkeeping, the entry is the record of an asset or a liability without a counterpart to that record (Ijiri 1986, 746). In "modern" single-entry bookkeeping, it is the record of income or expenses, also without a counterpart (IRS 2015). In double-entry bookkeeping, an entry is a debit or a credit record (IRS 2015).[7] In Boyle and Grigg's TEB, instead, the entries are the three *signature* records: the three parties (Grigg 2005). Furthermore, the single copy of the triple-signed record is in three places (ibid). This means that TEB does not challenge double-entry's bilateral recording of *transactions*.

## Triple-Entry Accounting

Triple-entry *accounting* presupposes triple-entry *bookkeeping*, but also exceeds it. The difference between one and the other goes in parallel with the distinction between *bookkeeping* and *accounting* themselves. Bookkeeping is defined as recordkeeping, i.e. simply keeping a sequential (chronological) record of transactions, whereas accounting builds on top of bookkeeping to make that information flow into the decision-making areas of the firm, by means of systematizing, compiling, collating, synthesizing, processing, analyzing and/or auditing.

The position that bookkeeping and recordkeeping are synonymous (and to distinguish them from accounting in the above-mentioned manner) appears to be the majority position in the peer-reviewed literature (Rukhiran and Netinant 2018; Vollmer 2003, 357) and in many instruction manuals and handbooks (Chandler 1977, 109-110; Ge 2005, 3; Ginigoada Foundation 2017, 3; Peters-Richardson 2011, 7; Wild, Shaw, and Chiappetta 2011, 4). This mainstream position will be followed in the remainder. However, it is worth noticing that some authors believe that there is no difference between bookkeeping and accounting (Lomax 1918, 74) or adopt different definitions (Edwards 1960, 447).

Also, we observe that both double-entry *bookkeeping* and double-entry *accounting* are terms in usage. However, the two entries are fundamentally a trait of the *bookkeeping* system, not the accounting itself. It is nevertheless legitimate to speak of double-entry *accounting,* because the specific accounting edifice is determined by the bookkeeping technique. However, this nuance should be kept in mind. Similarly, there is nothing intrinsically "triple" about TEA, yet the underlying changes in TEB should impact the accounting method to an extent, thus justifying the term.

[7] For a discussion of the different views on the specificity of double-entry bookkeeping, see Goldberg (1965, 215-219).

TEA is thus TEB *with an accounting solution*. In other words, TEA includes a shared transaction repository with a "signature – signature – signature" structure, but it is not limited to just sequentially storing transactions. Rather, it *also* serves to classify and interpret them, facilitating decision-making, financial analysis and forecasting, tax planning and financial reporting.

As an example, let us imagine that Alice buys from Bob two bicycles in 2019 which are identical in every possible way: one in March for 70 USD and one in April for 80 USD. In January 2020, Alice sells one bicycle to Charlie for 100 USD. In order to record the transactions, Alice proposes transaction entries over a system managed by Ivan. Bob and Charlie accept, Ivan verifies the validity of the transactions and the transactions get entered in a shared record. This is triple-entry *bookkeeping*. However, the question arises: what was Alice's profit from the transaction with Charlie?

Alice can resort to a number of methods to calculate the profit from each transaction, such as last-in first-out or LIFO, first-in first-out or FIFO, and average cost or AVCO (Peters-Richardson 2011, 104-107; Wild et al. 2011, 234-236). The choice between these methods will determine, for example, whether Alice made a profit of 20 USD, 30 USD or 25 USD in her transaction with Charlie. Nevertheless, this choice is a matter of accounting, not one of bookkeeping. Thus, having a shared transaction repository (TEB) does not by itself answer this question.

Either the choice between methods of assigning costs is left to Alice, Bob and Charlie, or it is pre-determined in the accounting software that they use. In any case, it is an accounting decision, with nothing specifically "triple" about its nature. However, Ivan may choose to integrate an accounting module[8] to the TEB system used by Alice, Bob and Charlie. In other words, Ivan may offer them to purchase a subscription to his online ERP system or webledger,[9] such that the TEB transaction records are automatically entered into each party's webledger. This may also allow Alice, Bob and Charlie to publish financial reports or to be audited, possibly in real-time. This accounting module built on top of the TEB record is called triple-entry *accounting*.

[8] General Ledger for Reporting (Boyle 2001b).

[9] Multi-user accounting suite.

## The REA Framework

Resource-Event-Agent (REA) is a model for an Enterprise Information System (Geerts and McCarthy 2006)[10] originally conceived by William E. McCarthy (1982). It proposes semantic abstractions generalizing business events (Boyle 2000c). This "computer software model of real-world (…) activities" is an "ontology" or "semantic model": a "set of classes, relationships, and functions in a universe of discourse" (Haugen and McCarthy 2000). The REA model purports to replace the classical double-entry accounting system with an information system integrated to all functional areas of an enterprise, i.e. not just limited to the accounting department. Nevertheless, REA can still be used to support the reporting artifacts of double-entry accounting, such as balance sheets and income statements. (Gal and McCarthy 1986, McCarthy 2001).

William E. McCarthy (1982) conceived REA as a solution to the deficiencies in existing accounting models, including imperfect classification schemes, the lack of granularity in data and the lack of integration of the accounting system with non-accounting information systems in an enterprise. To solve this, he designed REA as a conceptual modelling tool for a centrally defined database containing atomic transaction records with all the relevant variables. At the core of this model lies the representation of transactions as business *events*, where the companies' *agents* exchange *resources*.

Originally, McCarthy and his followers were spelling out the conceptual framework for an integrated business information system for the various areas of a *single* company. Nonetheless, with the advent of the Internet, the REA model was extended to multiple business entities in a trading community, that is, to *inter-company accounting*.[11]

Typically, accounting requires to record intercompany transactions from the perspective of one of the parties (ISO/IEC 2015, 3). For instance, "the mirror image of every sale recorded into receivables by a seller, is also recorded as a purchase, into accounts payable by a buyer" (Boyle 2000f). As a consequence, accounting records are viewpoint-dependent. McCarthy, however, proposed a representation of real-world business events from an independent, inter-enterprise view (ISO/IEC 2015, vii) that, simultaneously, was able to support different views of itself. This proposal was designed with the ANSI/X3/SPARC architecture for a Database Management System in mind (McCarthy 1982, 557; see also Tsichritzis and Klug, 1978) and eventually resulted in the Open-edi Distributed Business Transaction Repository (OeDBTR) project led by the ISO/IEC JTC 1/SC 32/WG 1 eBusiness Working Group, which is discussed below.

[10] In earlier versions, it was defined as a "generalized accounting framework" or "accounting model" (McCarthy 1982, p. 554, 2001).

[11] See, for instance, Haugen and McCarthy (2000), extending the REA model to conceive a single source of truth for an event-driven generalized representation of material flows throughout supply chains and demand chains.

## Differences and Similarities Between REA, Triple-entry Systems and Blockchain

REA is a generalized framework establishing an ontology, whereas both TEA and TEB can be deemed implementations of REA (see above for a discussion of the differences between TEA and TEB). Nevertheless, when applied specifically to inter-entity transactions, REA offers a concept that is functionally equivalent to TEB: The Open-edi Distributed Business Transaction Repository (OeDBTR).

OeDBTR is the term assigned, within the REA ontology, to a single-entry system which tracks the immutable history of events triggering changes of state in multiple business entities, relying on the independent view of the transaction as a single source of truth and the open-edi standard for electronic data interchange described in ISO/IEC 15944-21 (McCarthy and Holman 2019, see also Holman 2019). TEB and the OeDBTR thus share a fundamental characteristic: a viewpoint-independent record of transactions that is shared between two or more parties, and that can support different local "views" of the transactions.

How each transaction entry in a shared record manifests itself differently from the viewpoint of each party to the system is best represented as a 3-dimensional bookkeeping grid, by which N sheets equals the N parties to the system.

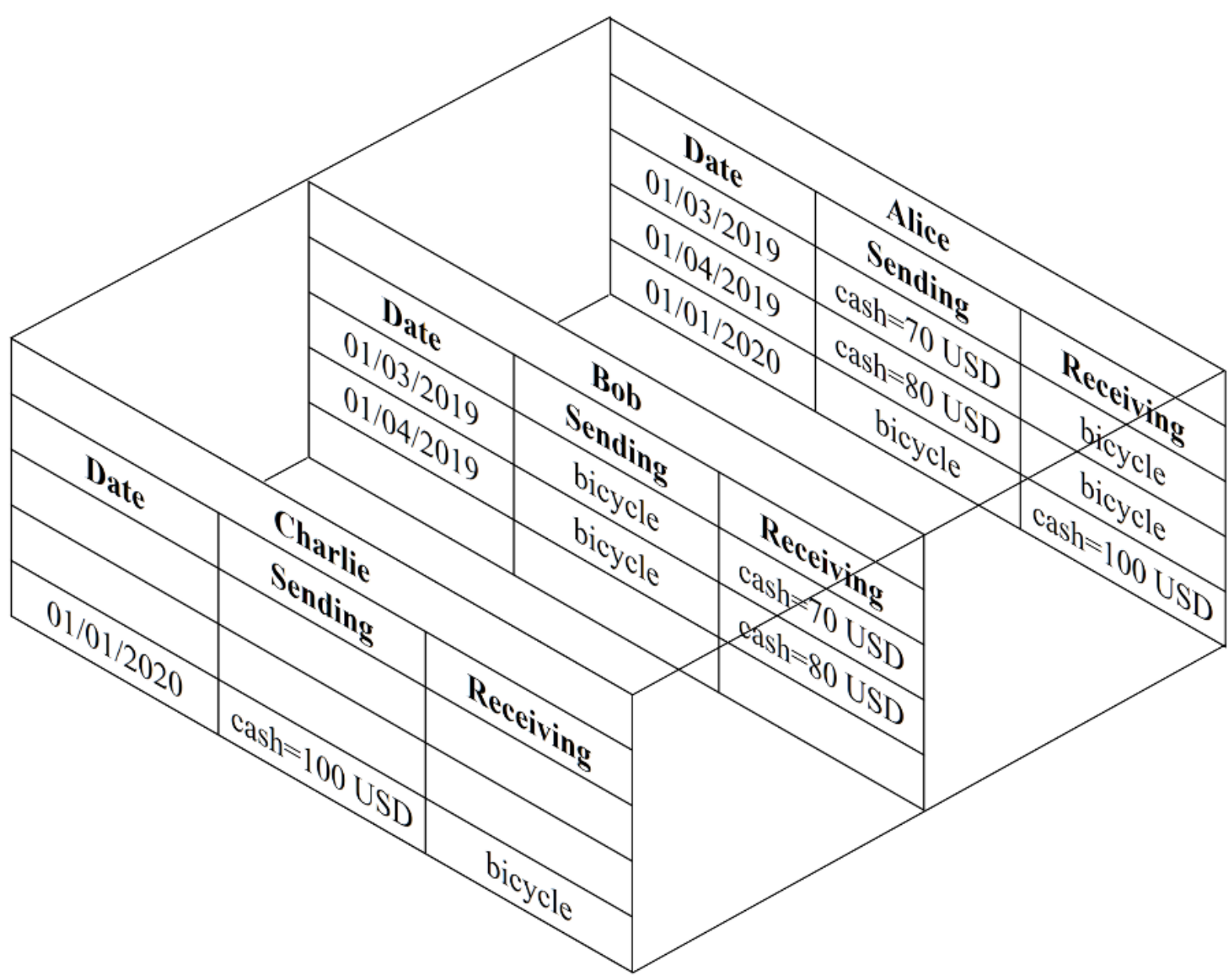

**FIGURE 2: A SIMPLIFIED THREE-DIMENSIONAL SHARED RECORD OF TRANSACTIONS. THE SAME TRANSACTION RECORD MANIFESTS DIFFERENTLY FROM THE VIEWPOINT OF EACH USER, I.E. IN EACH SHEET.[12] BASED ON BOYLE (2003A, 2003F).**

Historically, shared transaction records and three-dimensional accounting were technically infeasible. Therefore, each party to a transaction had to make its own, duplicated, viewpoint-dependent record of the same transaction in its own, two-dimensional books. We may call this “redundant bookkeeping.” However, as Internet-based shared data environments like triple entry systems or OeDBTRs becomes possible, redundancy can be eliminated. In consequence, we posit that TEB and REA’s OeDBTR are comparable.

Figure 3 presents a stylized comparison of the discussed accounting schemes. Both triple-entry systems and REA propose to eliminate redundancy through a shared transaction record constituting the single source of truth. REA provides for a shared “collaboration space” where the economic events in which agents participate are recorded (McCarthy 2016), but it does not specify the practical procedure necessary to agree on the single record. TEB introduces a signature-gathering process involving the two parties plus a trusted third party. In this context, TEB can be seen as a more concrete implementation, which takes this extra step at the cost of losing generalizability (Grigg 2020a).

[12] Boyle (2003a, 2003f) noted that it was possible to add more dimensions to the three-dimensional accounting grid, e.g. to allow a breakdown by type, month and/or purpose. In that context, the 3D accounting cube becomes a hypercube.

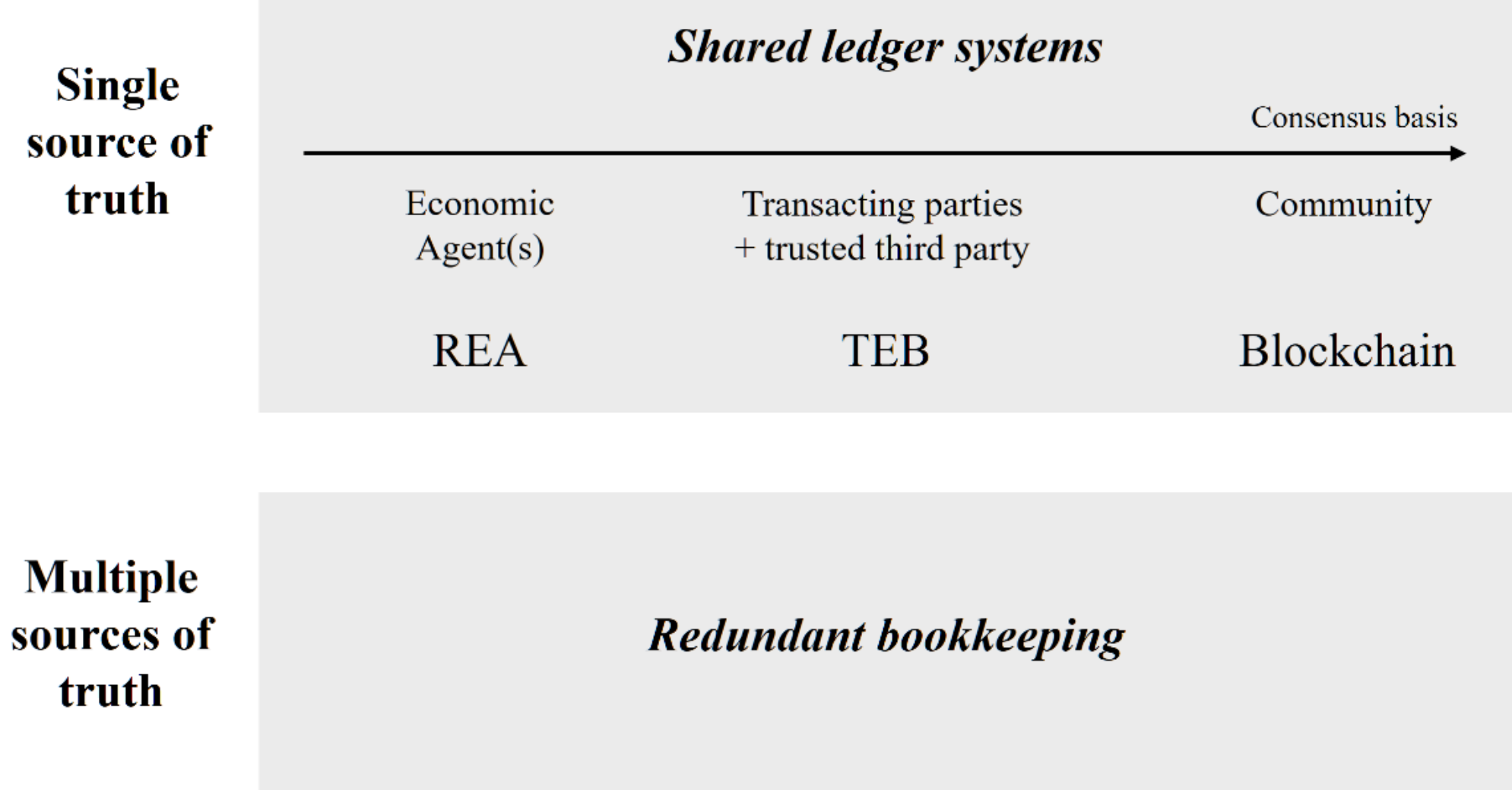

FIGURE 3: WITH TRADITIONAL REDUNDANT BOOKKEEPING, MULTIPLE SOURCES THAT DESCRIBE THE SAME ECONOMIC EVENT EXIST AND VARIOUS VERSIONS CAN ARISE. THIS DIFFERS FROM SHARED LEDGER SYSTEMS SUCH AS REA, TRIPLE-ENTRY BOOKKEEPING AND BLOCKCHAIN, WHICH INCORPORATE DESIGNS FOR A SINGLE SOURCE OF TRUTH. THE CONSENSUS BASIS ON WHICH THE TRUTH IS DETERMINED DEPENDS ON THE SPECIFIC SHARED LEDGER SYSTEM.

Blockchain technology develops the TEB model even further: it is a form of distributed software architecture that allows untrusted actors to securely agree on a transaction record without a central point of supervision (Tasca and Tessone 2019; see also Liu et al. 2019). To do so, it replaces the trusted third party that characterized original TEB designs with community-based consensus. In this manner, blockchain shows that the collaboration spaces envisioned in REA are practicable with current computational possibilities (McCarthy 2016).

## III. A GENEALOGY OF SHARED LEDGER SYSTEMS

As we described in Introduction, the *vox populi* history of shared ledger systems has so many gaps that its development is not just incomplete, but also improperly understood.

The popular version of the story goes as follows: Pacioli invented double-entry accounting in 1494. The convention remained unchallenged until 1982, when Yuji Ijiri conceived TEA (Vijai et al. 2019). Ijiri's ideas were forgotten, however, until Ian Grigg brought them back to life in 2005, giving a series of twists to the concept (ibid; Fullana and Ruiz 2020). In spite of this innovation, Grigg's idea was impracticable at the time because it was necessary to *trust* a third party with the shared ledger. Yet thanks to the fortunate exogenous appearance of Satoshi Nakamoto's Bitcoin whitepaper in 2008, suddenly it was possible to implement TEA without what impeded its viability: the need for trust (Cai 2019; Rao 2020).

But this story suffers from a number of problems. Notably, it overlooks the role of Todd Boyle in authoring the concept of triple-entry accounting. In consequence, the impact of the ideas of William E. McCarthy and Robert Haugen in Boyle is overlooked too (without prejudice to the originality of Boyle's work). Hence, the influence of the REA model in the genesis of TEA becomes completely excluded from the genealogical picture.

Furthermore, while Grigg's work was documented in 2005, much of it was undertaken between 1995 and 1997. Moreover, Ijiri's momentum accounting bore almost no relationship with Grigg's TEA. While a number of authors do point this out (Cai 2019; Dai and Vasarhelyi 2017; Gröblacher and Mizdraković 2019; Pacio 2018a; Wang and Kogan 2018), others appear to be unaware of this (Faccia and Mosco 2019; Faccia and Mostenau 2019; Vijai et al. 2019; Jeffries 2020). Note that Ijiri's exposition of accounting concepts did have a small influence on McCarthy. Although this can be seen as a historical connection between Ijiri and Grigg which is not recognized in the TEA literature reviewed, it is a minor one.

Finally, blockchain may not have been introduced as a completely exogenous invention that eventually enabled TEA. Instead, the opposite may be true. The Bitcoin blockchain accounting model has likely been influenced *by* TEA: Nakamoto apparently implemented Boyle's shared transaction repository idea, together with many other ideas discussed throughout the 1990s.

This section expands on the previous corrections,[13] so as to "set the record straight" and adequately conduct a genealogy of shared ledger systems.

## Early Antecedents to Triple-Entry Accounting: Momentum Accounting and the Resource-Event-Agent Model

The year 1982 brought forth the first major innovative challenges to the accounting status quo. A Japanese and an American accounting professor produced groundbreaking papers: Yuji Ijiri and William E. McCarthy, respectively.

[13] Also note that, while Pacioli (1494) did popularize double-entry accounting, Benedetto Cotrugli (1573) and Marino de Raphaeli (1475) had preceded him in beginning to introduce and develop the concept (Postma and Helm 2000; Sangster 2015; Sangster and Rossi 2018). Furthermore, comparable double-entry systems had been developed separately by the Italians, Koreans, and the second Muslim Caliphate at different times for the same purpose (Byeongju 2018; El-Halaby and Hussainey 2016; Zaid 2004).

Ijiri (1982, 1986, 1989, see also Hsieh 2018) explained that, in single-entry bookkeeping, only wealth (assets and liabilities) was recorded. The double-entry system incorporated income (flow accounts, revenues and expenses) so that, generally, one year's income statements explained the difference between two consecutive years' wealth statements:[14] the rate of change in wealth or "momentum"[15] (Ijiri 1986, 747). Thus, a third-entry to explain the rate of change in *income* would constitute a logical extension, that would make accounting systems "more dynamic and not focused on the present state (Balance Sheet), but on the future forecast" (Gröblacher and Mizdraković 2019, 60).

In other words, Ijiri had envisioned that a third entry would be used to explain the change between the income statements of two consecutive years, i.e. the rate of change of income or "force." Force, which could also be described as the *rate of change in the rate of change in wealth*,[16] is recorded in a third column named "trebit" (with debit ≡ credit ≡ trebit) together with wealth and momentum. In contrast, momentum and income are recorded in the "credit" column and assets are recorded in the "debit" column (Ijiri 1986, 751). Ijiri named this system "triple-entry bookkeeping," though it is also known as "momentum accounting."[17]

Almost simultaneously, McCarthy extended his earlier work (McCarthy 1979, 1980, 1981) on entity-relationship modelling (Dunn, Gerard, and Grabski 2016), and proposed an accounting framework for a shared data environment (McCarthy 1982, 554). McCarthy argued that a certain conceptual modeling would be "better able to support multiple "views" [multiple users] of a *centrally* defined database" (McCarthy, 1982, 555, italics are ours). This framework was at the antipodes of Ijiri's project to extend double-entry accounting principles, but it nonetheless drew some ideas from Ijiri's previous work on accounting measurements (see Ijiri 1975).[18] However, Ijiri's momentum accounting bore no relation with it.

The system would not have debits, credits, and accounts, which McCarthy deemed inessential for an accounting system, but mere "mechanisms for manually storing and transmitting data" (ibid, 560). Instead, McCarthy argued for recording detailed *transaction* histories which may be viewed by different (classes of) users (ibid, 569).

---

[14] Setting aside changes in equity for the sake of simplicity (Ijiri 1986, 747).

[15] Which becomes "income" when multiplied by duration (Ijiri 1986, 747).

[16] And becomes "impulse" when multiplied by duration (Ijiri 1986, 747-748a).

[17] Ijiri also advocated cryptosystem solutions involving public encryption keys and private decryption keys to protect business confidentiality (Ijiri and Kelly 1980, 118-120).

[18] While McCarthy (1982, 556) states that "the REA framework (…) is explained using the ideas of a number of accounting theorists, principally Yuji Ijiri," quotes concepts, uses ideas and expresses accounting principles following Ijiri (1975) on several occasions, Holman (personal communication, February 21, 2020) makes clear that "only a few vocabulary terms were adopted in the interest of consistent terminology, and nothing more." Ijiri (1993) defended the "beauty of double-entry bookkeeping," which McCarthy radically opposed. Thus, connections between the two authors can only be properly drawn if adequately contextualized, so as not to present them as allied researchers.

McCarthy also criticized traditional double-entry systems for, among other issues, the lack of granularity (excessive aggregation) in the data stored. REA accounting would fix this, be more efficient and accurate, and identify the agents involved, as well as other details, while preserving the duality (causal relationship) of economic events (Dunn, Gerard, and Grabski 2016, 554-555, 561). A more recent update to this criticism also included a critique for the lack of automation (McCarthy 2016).

Other precedents to shared ledger systems had a minor historical influence, but should still be mentioned. Much of the work that came out from the blockchain community in the last years is related to the cypherpunk movement of the late 20th century. Concepts discussed in mailing lists back then may have had an impact in the following decades, but a precise genealogy is hard to trace. In this context, Eric Hughes' open book accounting concept deserves a mention.

Back in 1993, Hughes (1993a) proposed a public transaction record with encrypted private balances. This would be a "single entry account" in a "shared funds account" that can moreover be expressed as double-entry bookkeeping for the parties. The accounts would be kept in accordance with each other through a public verification method (Hughes 1993b). Hughes, however, could not make this idea work technically (Grigg 2018).

## The Single Truth "Revolution"

> "We're followers of McCarthy's economic ontology, and ISO 15944-4." –
> Todd Boyle (2003c).

The 1990s and subsequent decades brought five important developments for shared ledger systems: Todd Boyle's coinage of the term TEA in its "modern" sense, Ian Grigg's cryptographic implementation thereof, Satoshi Nakamoto's Bitcoin whitepaper, McCarthy and Holman's participation in the REA-related ISO/IEC standards, and the emergence of actual TEA use cases.

In 1997, American accountant Todd Boyle (2001d, 2001e) moved back from Japan and set up General Ledger Dialtone in Seattle, an accounting solutions company specialized in webledgers While Boyle had come up *independently* with an idea of shared ledgers, he was later influenced by one of McCarthy's collaborators, Robert Haugen. Haugen was a software developer for a Core Components ebXML standards team (Boyle 2015 in Grigg 2014), who had worked in applying McCarthy's REA to supply chain Internet-based collaboration (Haugen and McCarthy 2000). Haugen introduced Boyle to McCarthy's work first, and McCarthy himself later, which had an impact in Boyle's ideas, consolidating his proposal for a shared ledger system.

Boyle (2000c) believed that McCarthy's REA framework was "high level (...) ahead of its time (...) [and] a goldmine" with many merits,[19] but envisioned a much more valuable application for it: TEA. This would be commercially launched as an Internet-based, multi-company, and low-cost accounting software. It would enable large supply chains (value networks) supported through a webledger "spitting out" the transactions therein (ibid), i.e. a "general ledger (...) where independent companies could post their resource transfers" (Boyle 2015 in Grigg 2014; see also Haugen and McCarthy 2000). The webledger could contain all internal general ledger entries but its principal use was to "contain those journal rows involving external parties" (Boyle 2001f). To enable automatic reconciliation and external (as well as internal) integration, inter-ledger semantics based on the GAAP would have been developed to allow different business systems to interface and form a coherent whole (Boyle 2002).

This was designed to have a classic double-entry accounting structure but a non-double-entry interface, for user friendliness (Boyle 2000a), which reflects another of Boyle's criticisms of REA: he thought that, even though REA was superior to double-entry, the quest to replace the latter with the former was "a distraction," as double-entry "merely" records data alongside a business system (Boyle 2000c; see also Boyle 2000a).

For this purpose, as well as for other commercial transactions, Boyle believed that a mechanism to communicate both parties in a transaction where and when an economic exchange has happened ("recognition") should be built in a joint web accounting application (a B2B middleware server; Sachs 2001 in Boyle 2001c; Boyle 2001f), rather than delegated to each participant (Boyle 2001c). The mechanism should act as an encrypted "public document repository service," as a "notification service," as a service to record replies (e.g. acceptances), and as an archive and reporting service to provide "persistent and responsive storage of inter-party transactions, sufficient to achieve a robust and intrinsic reconciliation" (Boyle 2001f).

Boyle's webledger architectures would implement this solution, in the form of a "shared" or "public transaction repository" (STR or PTR) based on "single-entry hosted transaction tables" (Boyle 2000e, 2000d, 2001a). There would thus be a single, shared, network-centric record but, because of the triple-signed structure, the system would be called triple-entry accounting (Boyle, 2001b). Boyle (2000c) proposed REA to support the model for the back-end software, and even developed an REA-based economic ontology to conceptually describe this system (Boyle 2003d).

[19] Boyle also criticized REA for being mislabeled as an accounting system (when, in his view, it was really a very generalized business information system). He moreover believed that the circumstances that REA had originally come to solve in the 1980s did not exist anymore (Boyle 2000c). For a response, see Haugen (2001) and McCarthy (2001).

> "Thanks to Bill McCarthy and his REA school, who were the source of most of these ideas" – Todd Boyle (2003d).

At approximately the same time that Boyle developed GL Dialtone and TEA, oil markets consultant and researcher Chris Cook independently developed OilClear, a petroleum-specific STR concept. Cook (2002) argued that "a 'shared transaction repository' and a 'shared title repository' (...) connected by clearing and settlement software" were necessary for the new market structure in the age of the Internet and instantaneous communication. The concept, called "Market 3.0," reportedly hit the "Internet neutrality-liquidity paradox," did not find a route to the market and was later fragmented and appropriated by ICE's eConfirm and by CME's Tradehub. While ICE's eConfirm and by CME's Tradehub did not stop resorting to exchanges (Cook 2016), OilClear set a milestone nonetheless.

Also independently from Boyle and Cook (Grigg 2019b), Ian Grigg together with Gary Howland co-developed a similar concept between 1995 and 1997: the Ricardo payment system and Ricardian contracts, documented in Howland (1996).[20] This idea, which Grigg (2000, 2004, 2005) kept developing afterwards, was an attempt to replicate how economic events are recorded internally *within* firms through an ERP system, in a shared data environment *between* firms. It involved a shared set of receipts for the transactions common to two parties, and a trusted third party limited to signing, timestamping and ordering. While, originally, two different receipts were conceived: one for the payer and one for the payee (Howland 1996), shortly after this idea was abandoned, folding the two receipts into a single one. This constituted the genesis of triple-entry.[21]

This shared receipt would be the *dominating* record for a transaction. Moreover, in Grigg's design, the receipt is not just evidence for the transaction: it *is* the transaction itself, because it holds all the relevant information so as to build an entire data processing concept around it.[22] Furthermore, in order to prevent any disputes around semantics, these are locked down through a Ricardian Contract (Grigg 2005b): a human and machine-readable text file containing both the terms of an agreement and the program executing the financial instrument, such that they are the same thing, i.e. "the issue is a contract" (Grigg 2004, 2000). The parties hold a (cryptographic) key to authorize each transaction and a copy of the receipts issued by the accounting agent (see also Boyle 2003b). To modify the record, the accounting agent needs the signature of both parties. In other words, every modification of the record requires a three-party consensus.

[20] Ian Grigg was not explicitly credited in this paper. However, the paper states that the company founded by Grigg (Systemics Ltd.) developed the system, and Grigg himself reports being a co-developer in Grigg (2000).

[21] Grigg, I. (personal communication, April 15, 2020).

[22] For example, it is possible to calculate Bob's balance by reading and calculating the net sum of all the receipts mentioning Bob.

## The Single Truth Convergence

In 2000, Grigg (2000) began to document his work. In 2004, upon realizing that his design could have radical implications for accounting, Grigg pursued further development of his ideas (ibid), which he called triple-entry *accounting*, but later pointed out that were really triple-entry *bookkeeping* (Grigg 2019a). A draft of the resulting paper was circulated in June 2005, with Boyle commenting on it. Boyle noted that he had been working on the same idea for years as well (Grigg 2016a). As a consequence, Grigg integrated and implemented many of Boyle's ideas within the paper. However, while a draft of the paper "credited Todd Boyle as an author, (...) this was later withdrawn at his request due to wider differences between the views" (Grigg 2005b). These differences were related to the breadth of the scope or generalizability of the model.

Grigg first became aware of McCarthy's REA concept in 2017 through Boyle introduced (Grigg 2017a), but McCarthy's influence was present nonetheless. He was also unaware of Ijiri's work (Grigg 2020), though his imprint on TEA had certainly been almost trivial.

Note that, according to Grigg (2017a, 2017b, 2017c, 2020a), his *TEB inadvertently implemented key ideas more generally contained in REA*: "the Receipt as I describe it in the paper and as it is used, is an REA construct converted to data; the (hash of the) Ricardian contract is the resource, the signing/timestamping by Ivan is the event, and the payee/payer are the agents."[23]

However, before the advent of blockchain, a workable triple entry system would have necessitated a trusted third-party intermediary, who would also have been susceptible to attack, error, or loss. The intermediary would have been vulnerable, like the transacting parties themselves. The invention of Bitcoin and its blockchain permitted an adaption of Grigg's theory without a single center (Grigg 2019c).

The new solution consisted of replacing the central intermediary with a decentralized ledger, to which the customer's and supplier's accounts are connected, in which both sides of a transaction are recorded, and thus having the entries reaching a consensus. This application may be Business-to-Business (B2B) or Government-to-Government (G2G), e.g. between companies' tax and royalty payments to governments.

[23] Grigg, I. (personal communication, April 15, 2020). For further discussion on how TEA may offer an implementation of more generalized ideas contained in REA, see Grigg (2020b).

However, blockchain might have been an *endogenous* technological development that enabled TEA. Firstly, there are common sources to the ideas of both Boyle and Nakamoto.[24] Furthermore, there are remarkable architectural similarities between Boyle's ideas and Bitcoin: Bitcoin is a pseudonymous, immutable public transaction repository with an integrated payment layer underpinned by a triple-entry structure (in which the trusted third party is the distributed ledger or community). Moreover, there are anecdotal reasons to think that Boyle's idea of a publicly shared transaction repository was one of many sources in the corpus of preceding work on top of which the Bitcoin edifice was built.

In this vein, one ought to point out that Grigg (2014) has stated that Boyle's "notion of a public, and/or shared ledger is one of those components employed in Bitcoin (...) Satoshi Nakamoto stands on the shoulders of giants, his design is the very clever assembling of components that were tried beforehand." Note that Grigg himself is closely associated with Satoshi Nakamoto.[25] The value of Boyle's ideas has furthermore been considered a precursor to blockchain in other discussions within the cryptographic community (Brown and Grigg in Brown 2015; Grigg in Swanson 2015; Sleeter 2014). Nonetheless, since Nakamoto's identity has not been fully confirmed (and might never be), this evidence is limited.

As stated above, Grigg was unaware of McCarthy's REA model and his influence over Boyle's work. Considering also that most TEA use cases and papers follow Grigg's idea (not Boyle's), it is therefore unsurprising that recent developments in the REA world have remained unnoticed despite their high pertinence to TEA. Nevertheless, these developments keep happening, with the work of ISO/IEC JTC 1/SC 32/WG 1 – a working group of a subcommittee of the joint committee between ISO and IEC – being most relevant.

The ISO/IEC 15944-4 standard was published in 2007,[26] then updated in 2015 and is currently under review. It uses the REA ontology to model a formal framework for business transactions named "Open-edi Business Transaction Ontology (OeBTO)" (ISO/IEC 2015, v; see also Dunn et al. 2016, 555). This framework maintains that the redundancy in mirroring records of a transaction must be abandoned to eliminate the possibility of inconsistencies (ibid, vi) and because it is viewpoint-dependent. In turn, it proposes an independent, inter-enterprise view of transactions.

---

[24] Boyle was a part of the same cypherpunk mailing lists in which cited influences in the Bitcoin whitepaper like Adam Back and Wei Dai participated (see Nakamoto 2008; Venona Cypherpunks Archives 2004).

[25] Some believe Grigg *is* Satoshi Nakamoto himself, based on, inter alia, stylometric studies (Bits n Coins, 2019; Helsel 2018), a claim we are unable to verify. Grigg (2016c) has denied being a member of the Satoshi Nakamoto team. However, Grigg (2016b) has also claimed direct knowledge of its internal workings.

[26] Boyle (2003c) claimed to be a follower of ISO 15944-4 in 2003, which means that he was aware of the draft before publication.

While ISO/IEC 15944-4 mostly provides definitions, the joint work of Holman and McCarthy (2019, see also Holman 2019) built upon it in designing ISO/IEC 15944-21, a standard providing guidance on the implementation of an OeDBTR, i.e. a (typically, but not necessarily, blockchained) shared transaction repository remarkably similar to a TEA system, within the REA ontology. A draft of this standard has already been registered and approved, but the publication process has yet to be completed (ISO 2019).

This has interesting implications, as it opens the door for TEA systems to follow ISO/IEC specifications. In fact, one of the TEA use cases listed in Figure 1, bBiller, is an OeDBTR implementing the REA ontology. Another TEA use case, Pacio, incorporates the REA ontology in the Standardised Semantic Information Model Database of Facts in the TEA and IDEA diagrams of its whitepaper (Pacio 2020, 6, 9). This facilitates the possibility of a TEA-REA reconciliation, in spite of the neglected influence of REA in TEA.

Moreover, further convergence is conceivable. We established that Yuji Ijiri did not influence TEA, except for the adoption by McCarthy of the terminology laid out in Ijiri's works prior to momentum accounting. Nevertheless, Ian Grigg (2020b) recently stated that there is a connection between TEA and Ijiri's *momentum accounting* and, furthermore, a potential for symbiosis: Ijiri's momentum accounting requires complicated calculations, but these calculations cannot be reliable if the underlying records are subject to error. This made Ijiri's model too risky for the market (ibid). The execution of momentum accounting on top of cryptographic triple-signed receipts, however, might allow the former to perform reliably.

## IV. DISCUSSION

Shared ledger systems are one of the most important innovations of the past decades. They are not a panacea, not replacing many of the traditional functions of accounting and not sufficing to prevent fraud, money laundering, etc. by themselves. Nevertheless, shared ledger systems have led the way for accounting applications in the Internet era. In particular, TEA is one of the pioneering concepts for accounting in shared data environments, delivering many benefits by enabling or facilitating external integration, instantaneous reconciliation, lower redundancy, low-cost real-time auditing, financial reporting, invoice automation, dispute resolution, etc. (Alawadhi, et al. 2015; Boyle 2002, 2003f; Dai 2017; Dai and Vasarhelyi 2017; ICAEW 2018; Mohanty 2018, 47; Request 2018a, 2018b)

The lack of an integral genealogy of TEA has obscured the role of the *accounting discipline* in giving birth to TEA. Specifically, Boyle's work was overlooked, and so was the influence of the REA model over TEA. Consequently, the point that TEA is to some extent a historical byproduct of McCarthy's research was rarely raised in public discourse. As a consequence, REA and TEA have remained two separate streams of research. Yet it would be very important to bring these streams together.

Furthermore, since Satoshi Nakamoto's Bitcoin may likely have been inspired by Boyle's TEA (as well as other influences), McCarthy's REA model may have had an indirect historical impact on the genesis of the blockchain technology itself. It is often said that "blockchain is fundamentally an accounting technology" (ICAEW 2018, 1). In the light of these findings, this statement may be truer than ever thought before.

Indeed, the structural resemblances between REA and TEA are not just a coincidence, but a natural outcome given the historical influence of the former over the latter. This may explain why bBiller, for instance, is considered a TEA use case (Pacio 2018b), but their designers consider it a REA use case. It may also explain why Pacio incorporates part of the REA ontology in its TEA design. Furthermore, if TEA was indeed an influence for blockchain technology, one would similarly expect subsequent blockchain projects to intuitively use TEA principles in their solutions without deliberately implementing the idea nor necessarily using the term.

## V. CONCLUSION

In this paper, we attempt to trace three intersecting development pathways that represent incarnations of shared ledger systems. In particular, we explore possible connections among the long-established REA and TEA frameworks, and the nascent blockchain technology. By filling in the gaps in the genealogy of shared ledger systems, we correct historical misconceptions, and give due credit to related prior works that have been insufficiently recognized. A clearer understanding of the historical evolution of shared ledger systems potentiates the academic debate, as well as further discourse among researchers in Resource-Event-Agents, triple-entry accounting and blockchain.

## APPENDIX A: ACRONYMS

| ANSI/X3 | American National Standards Committee on Computers and Information Processing |
|---|---|
| AVCO | Average Cost |
| B2B | Business-to-business |
| DLT | Distributed Ledger Technology |
| FIFO | First-in, First Out |
| G2G | Government to Government |
| GAAP | Generally Accepted Accounting Principles |
| GL | General Ledgers |
| IEC | International Electrotechnical Commission |
| ISO | International Organization for Standardization |
| LIFO | Last-in, First-Out |
| OeBTO | Open-edi business transaction ontology |
| OeDBTR | Open-edi Distributed Business Transaction Repository |
| PTR | Public Transaction Repository |
| REA | Resource-Event-Agent |
| SPARC | Standards Planning and Requirements Committee |
| STR | Shared Transaction Repository |
| TEA | Triple-entry accounting |
| TEB | Triple-entry bookkeeping |

## APPENDIX B: ACKNOWLEDGEMENTS

We thank the following individuals for answering our questions per email correspondence, per social networks or per interview:

| Individual | Position |
|---|---|
| Chris Cook | Senior Research Fellow at UCL |
| Chris Odom | Founder of Open-Transactions |
| Craig S. Wright | Chief Scientist at nChain |
| David Hartley | CEO of Pacio |
| G. Ken Holman | CTO at Crane Softwrights Ltd and editor of ISO/IEC 15944-21 |
| Ian Grigg | CTO at Solidus |
| Robert Haugen | Developer at Mikorizal Software |
| Todd Boyle | Founder of GL Dialtone |

Commenting, however, does not equate to endorsement of the views expressed in this paper.